# Discovery of a new nontoxic cuprate superconducting system Ga-Ba-Ca-Cu-O

Yue Zhang, Wenhao Liu, Jin Si, Xiyu Zhu*, Chengping He, Haonan Zhao, and Hai-Hu Wen*

Center for Superconducting Physics and Materials,
National Laboratory of Solid State Microstructures and Department of Physics,
National Collaborative Innovation Center of Advanced Microstructures, Nanjing University, Nanjing 210093, China

**Abstract:** Superconductivity is observed in a new nontoxic cuprate system Ga-Ba-Ca-Cu-O, with $T_c$ = 82K for $GaBa_2Ca_5Cu_6O_{14+\delta}$ (Ga-1256) and $T_c$ = 116K probably for $GaBa_2Ca_3Cu_4O_{10+\delta}$ (Ga-1234) or $GaBa_2Ca_2Cu_3O_{8+\delta}$ (Ga-1223), respectively. All compounds are fabricated by solid state reaction method under high pressure and high temperature. Samples are characterized by resistivity, magnetization and X-ray diffraction (XRD) measurements. The temperature dependence of magnetization measured in both zero-field-cooled and field-cooled processes on one sample (S1) shows two superconducting transitions at about 82K and 113K. The estimated shielding fraction for the phase with $T_c$ of 82K is about 67%, while the fraction for another phase with $T_c$ of 113K is quite small. The XRD Rietveld refinement for S1 indicates two main phases existing in the sample, Ga-1256 with fraction of about 58% and non-superconducting $Ca_{0.85}CuO_2$ with fraction of about 42%, respectively. Thus, we can conclude the superconducting phase with transition temperature of 82K is due to Ga-1256. The resistivity measurement also confirms the superconductivity for S1, and the resistivity reaches zero at about 82K. The temperature dependence of magnetization for another sample (S2) shows much higher superconducting shielding fraction for the phase with $T_c$ of 116K, which may be a promising prospective for the synthesis of Ga-1234 or Ga-1223 phase.

**Keywords**: cuprates, high pressure and high temperature, superconductivity

**PACS number(s)**: 74.25.Fy, 62.50.-p, 74.25.Ha

## Introduction

Since the first discovery of high temperature superconductivity in LaBaCuO in 1986 [1], the research on cuprates have become a core issue in the field of superconductivity. This event was followed by very efficient efforts leading to the discovery of superconductors with transition temperature $T_c$ beyond the liquid nitrogen temperature in $YBa_2Cu_3O_{7-\delta}$ [2,3]. And since the discovery of Tl-based high-temperature cuprate superconductors with $T_c$ of about 120K [4], many other A-Ba-Ca-Cu-O systems such as Hg-Ba-Ca-Cu-O, Au-Ba-Ca-Cu-O, (Cu,C)-Ba-Ca-Cu-O have been discovered or synthesized. Among them, the Hg-based (Hg-1223) system holds the record of the highest superconducting transition temperature of 133K at ambient pressure [5], and up to 164K under high pressure [6]. Although the Hg-based and Tl-based systems have rather high transition temperatures, the toxic elements Hg and Tl strongly limit the application of these materials. Among the nontoxic systems in cuprate family, such as Y-based $YBa_2Cu_3O_{7-\delta}$ (Y-123), Bi-based $Bi_2Sr_2CaCu_2O_{8+\delta}$ and $Bi_2Sr_2Ca_2Cu_3O_{10+\delta}$ (Bi-2212 and 2223) and $(Cu,C)Ba_2Ca_{n-1}Cu_nO_y$ systems, the Bi-based system also has critical temperature above 100K, and it shows excellent critical current density at 4.2K with magnetic field higher than 20T, so it is promising for application with high magnetic field at liquid helium temperature [7,8]. However, in the liquid nitrogen temperature region, the irreversibility field and critical current density are strongly suppressed for the Bi-based systems Bi-2212 [9] and Bi-2223 [10]. Thus, the Y-based $YBa_2Cu_3O_{7-\delta}$ with $T_c$ of about 90K, as a nontoxic material with high irreversibility line [11,12], is still considered to be a promising material for application in the liquid nitrogen temperature region. Recently, our group finds that, in the nontoxic phase $(Cu,C)Ba_2Ca_3Cu_4O_{11+\delta}$ synthesized under high pressure and high temperature possesses not only high transition temperature at $T_c = 116K$, but also high irreversibility line and high critical current density $J_c$, which may be a good candidate for applications in the temperature region of 77K to 100K [13]. In this paper, we report the discovery of superconductivity in a new nontoxic cuprate superconducting system Ga-Ba-Ca-Cu-O with $T_c = 82K$ for Ga-1256 phase and $T_c = 116K$ probably for Ga-1234 or Ga-1223 phase. The structure and superconducting properties are characterized by X-ray diffraction (XRD), magnetization and resistivity measurements.

## Experimental details

The bulk Ga-based cuprate samples are synthesized by solid state reaction method under high pressure and high temperature using the precursors $BaCuO_{2.13}$ and $Ca_2CuO_3$. The $BaCuO_{2.13}$ is obtained by calcining $BaO_2$ (Aladdin, powder, purity 95%) and CuO (Alfa Aesar, powder, purity 99.995%) at 900°C for 60h in flowing oxygen atmosphere. $Ca_2CuO_3$ is prepared first by calcining a well-ground mixture of $CaCO_3$ (Alfa Aesar, powder, purity 99.99%) and CuO at 950°C for 20h in air. The obtained $Ca_2CuO_3$ is ground and then sintered in flowing oxygen gas at 950°C for 40h. Afterward, the compounds of $BaCuO_{2.13}$, $Ca_2CuO_3$, $Ga_2O_3$ (Aladdin, powder, purity 99.99%), CuO and if necessary, an appropriate amount of $Ag_2O$ (act as oxidizer) are thoroughly ground and mixed, and then pressed into a pellet and sealed into a gold capsule. For the final high pressure and high temperature synthesis, the pellet enclosed by the gold capsule is placed into a BN container, surrounded by graphite sleeve resistance heater and pressure transmitting MgO rods. The final reaction is carried out at 1150 to 1170°C under 3.7GPa for 1-5h, then cooled down to room temperature in 5 min before the pressure was released.

The X-ray diffraction (XRD) was measured using a Bruker D8 Advanced diffractometer with the CuK$_{\alpha 1}$ radiation at room temperature. The temperature dependence of dc magnetic susceptibility was measured by a Quantum Design instrument SQUID-VSM7T. The measurement of magnetization hysteresis loops was performed with magnetic fields up to 7T. We also measured the resistivity versus temperature by the standard four-probe method using a physical property measurement system (PPMS16T, Quantum Design).

## Results and discussion

Fig. 1(a) shows the temperature dependent magnetic susceptibility of S1 measured in zero-field-cooled (ZFC) and field-cooled (FC) modes with a magnetic field of 10Oe. As shown in the inset, there are two superconducting transition temperatures at around 82K and 113K, respectively. Assuming an average mass density of 6 g/cm$^3$ for the sample, we have $4\pi\Delta\chi \approx$ -67% for the 82K superconducting phase determined by using the diamagnetic susceptibility measured with the ZFC mode at 10 K. Another inset shows an enlarged view in the temperature region of 100 to 125K, a superconducting transition is clearly visualized at about 113K. Combining the following results of XRD, we can conclude that the phase with $T_c$ = 82K is from Ga-1256. As the screening fraction for the 113K phase is too small, we can not attribute it to any specific compound at this moment. However, according to our knowledge of cuprate, the $T_c$ is very close to Tl-1234 and Tl-1223. Thus, the phase with $T_c$ = 113K probably comes from Ga-1234 or Ga-1223. Temperature dependence of resistivity under zero applied magnetic field is shown in Fig. 1(b), it shows a metallic behavior above $T_c$. Since we have two superconducting phases here, the transition temperature corresponding to the low temperature phase is hard to be determined from resistivity. By using a linear extrapolation of high temperature data of resistivity, we found the deviating point at about 113K which was determined as the transition temperature of the high-$T_c$ phase.

In Fig. 2, magnetization hysteresis loops (MHLs) of S1 are shown for temperatures of 10K and 40K with applied magnetic fields up to 7T. The sweep rate of field below 0.3T is 10Oe/s and that higher than 0.3T is as fast as 100Oe/s. We also measured the MHL at 130K which is slightly higher than $T_c$ of the high-$T_c$ phase and took it as the background. And then we deducted the background from the raw data of 10K and 40K. The magnetization hysteresis ΔM is small even at a low temperature of 10K, comparing with other systems such as Bi-2223 [14], Y-123 [12,15] and (Cu,C)-1234 [13,16] which may be caused by phase segregation and the poor connecting of the superconducting phase. As we can see, comparatively, the MHL at 10K is much wider than that measured at 40K, which supports the picture of phase segregation mentioned above.

Fig. 3(a) shows the X-ray diffraction pattern for the sample S1 and the Rietveld fitting curve (the red line) to the data [17]. The samples synthesized so far always contain the impurity phase $Ca_{0.85}CuO_2$. Thus as the first step, we deduct the XRD peaks of the phase $Ca_{0.85}CuO_2$, the left peaks are fitted by the Fullprof program with a self-consistent fitting to the peak positions [18], which gives the space group and lattice constants of P4/mmm symmetry with parameters a = 3.85Å and c = 25.33Å. This result is very close to the 1256 phase in cuprate [19]. Then we refine the whole XRD by the model of 1256 with the impurity phase $Ca_{0.85}CuO_2$ by Rietveld fitting. As we can see, the data can be fitted very well with R-weighted pattern Rwp = 5.57%, Rp = 4.28%. The refinement gives a main tetragonal Ga-1256 phase (taking a fraction of about 58%) of P4/mmm symmetry with parameters a = 3.85Å and c = 25.47Å. The main impurity phase $Ca_{0.85}CuO_2$ which marked by the vertical cyan bars takes a 42% fraction. The XRD result is consistent with the superconducting volume fraction determined by the magnetization measurement. Fig. 3(b) shows the schematic structure of Ga-1256 with the space group P4/mmm. As we can see, for one unit cell, the system has six layers of $CuO_2$ planes.

In Fig. 4, we demonstrate temperature dependent magnetic susceptibility of another sample (S2) with an external magnetic field of 10Oe. This sample (S2) was synthesized with the conditions similar to S1 but the sintering time is only 1hour which is shorter than that of S1. In addition, the oxygen content of S2 is slightly less than that of S1. The superconducting critical temperature $T_c$ is about 116K which may correspond to Ga-1234 or 1223 phases. This is consistent with the result of the second superconducting step (at 113K) shown in Fig. 1, a slight change of $T_c$ may be caused by the variation of the oxygen content. Similarly, assuming an average density of 6 g/cm$^3$ for S2 and using the ZFC data at 10K, a calculation of the magnetic screening volume is only about 10% which is still quite small and more effort is needed to improve the sample quality for the high $T_c$ phase.

## Conclusions

In summary, we synthesized a new nontoxic cuprate system Ga-Ba-Ca-Cu-O by solid state reaction method under high pressure and high temperature. The XRD pattern of S1 shows a tetragonal Ga-1256 phase of P4/mmm symmetry with a main impurity $Ca_{0.85}CuO_2$. The DC magnetic susceptibility measurement displays two superconducting transition temperatures at around 82K and 113K, which may correspond to Ga-1256 and 1234 (or 1223) phases, respectively. Resistivity under zero magnetic field also confirms the superconductivity. The magnetization result for another sample (S2) shows a single transition temperature at 116K with a magnetic screening volume of 10%.

**Acknowledgements** This work was supported by the National Key Research and Development Program of China (Grant Nos. 2016YFA0300401, and 2016YFA0401704), and the National Natural Science Foundation of China (Grant Nos. A0402/11534005, and A0402/11674164).

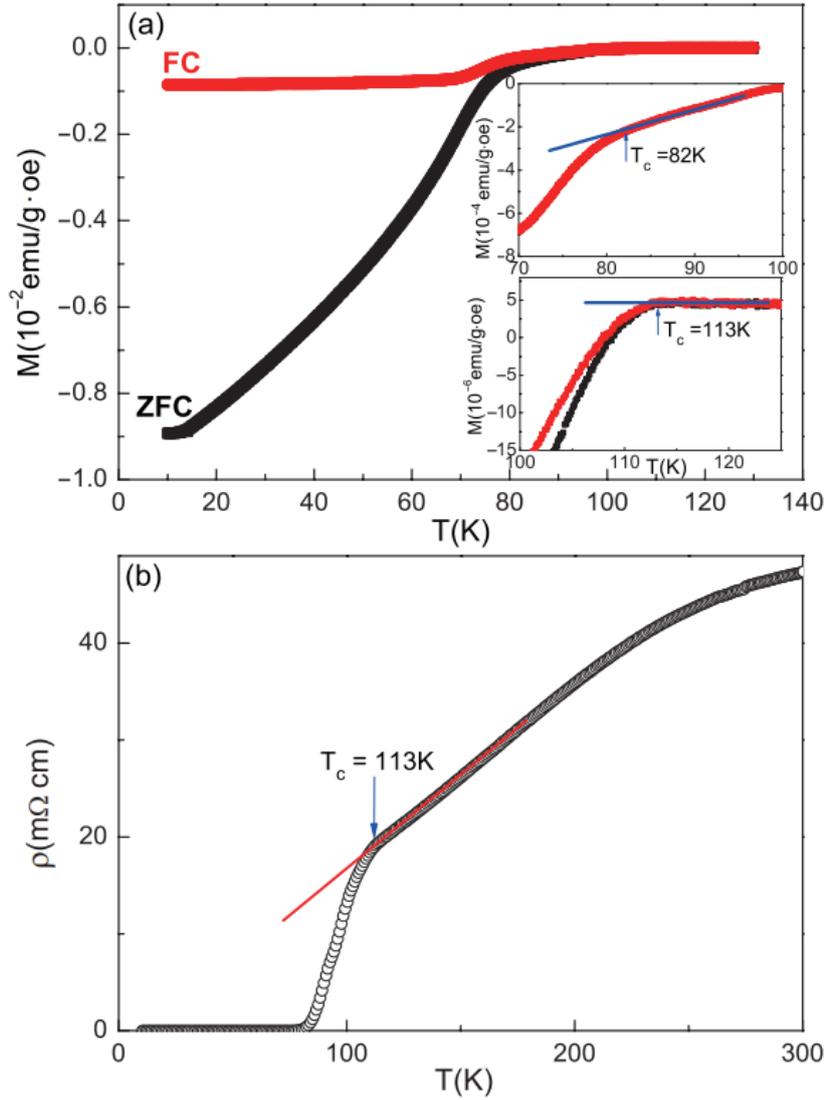

**Figure 1** (Color online) (a) Temperature dependence of magnetic susceptibility of S1 measured in ZFC and FC modes under a magnetic field of 10Oe. Inset shows the enlarged view for the phase with $T_c$ = 82 K in the temperature range of 70 to 100 K, and another inset for the phase with $T_c$ = 113 K in the temperature range of 100 to 125 K. (b) Temperature dependence of resistivity under zero applied magnetic field.

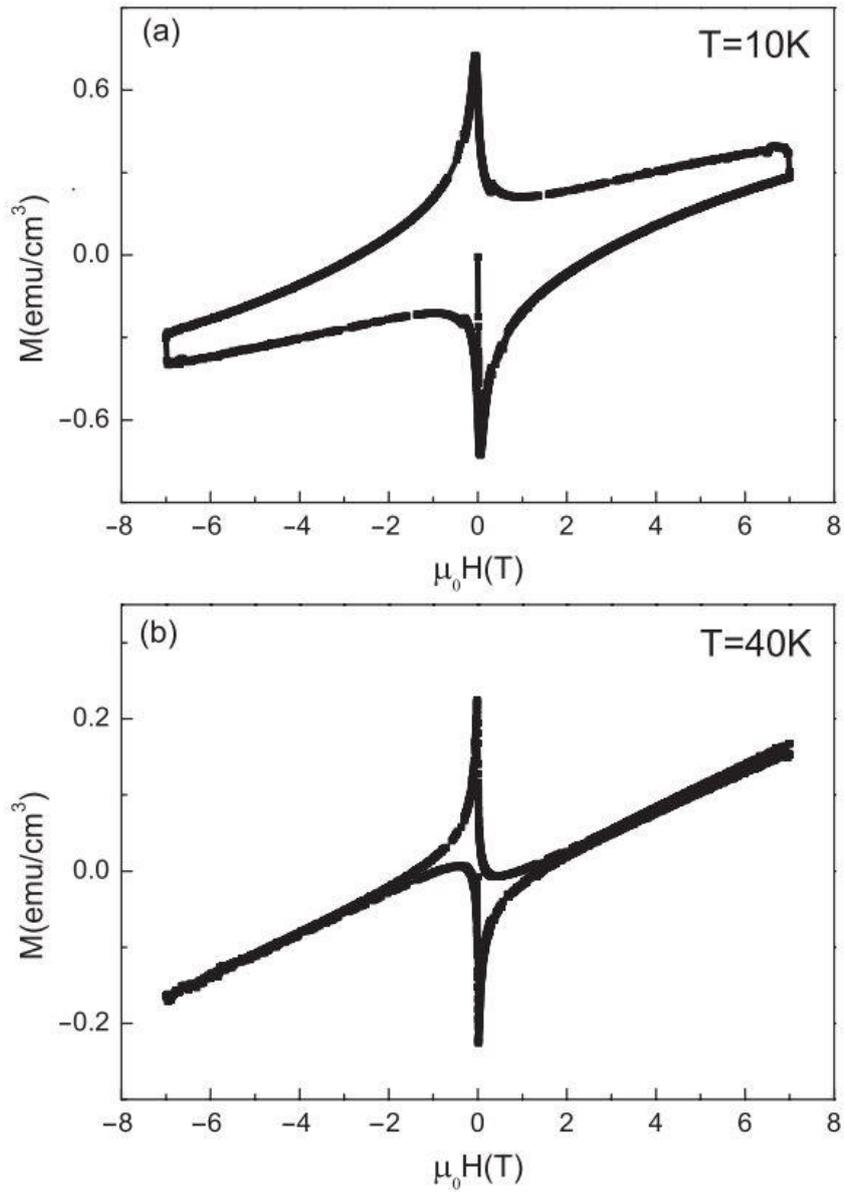

**Figure 2** (a) Magnetization hysteresis loop (MHL) of S1 at the temperature of 10 K with applied field up to 7 T. (b) MHL for the same sample at 40 K under the magnetic field up to 7 T. We have deducted the paramagnetic background measured at 130 K from the data.

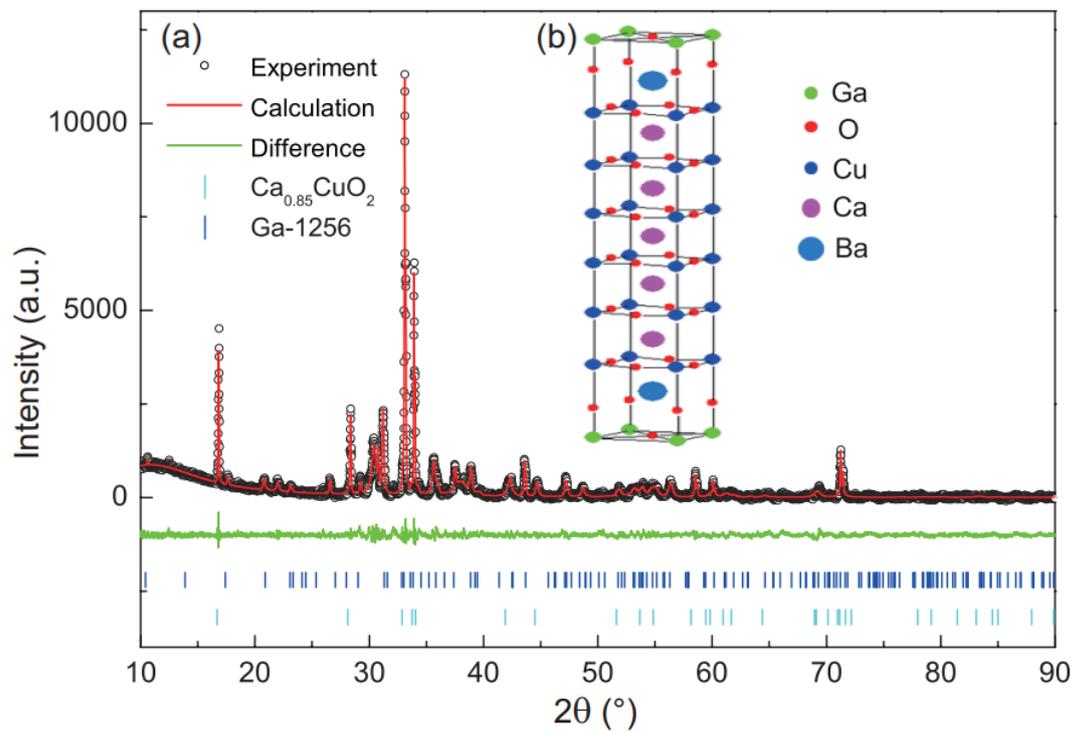

**Figure 3** (Color online) Powder XRD pattern and structure model for S1. Most of the peaks in XRD can be indexed with the main phase Ga-1256. The rest are indexed as the impurity phase $Ca_{0.85}CuO_2$.

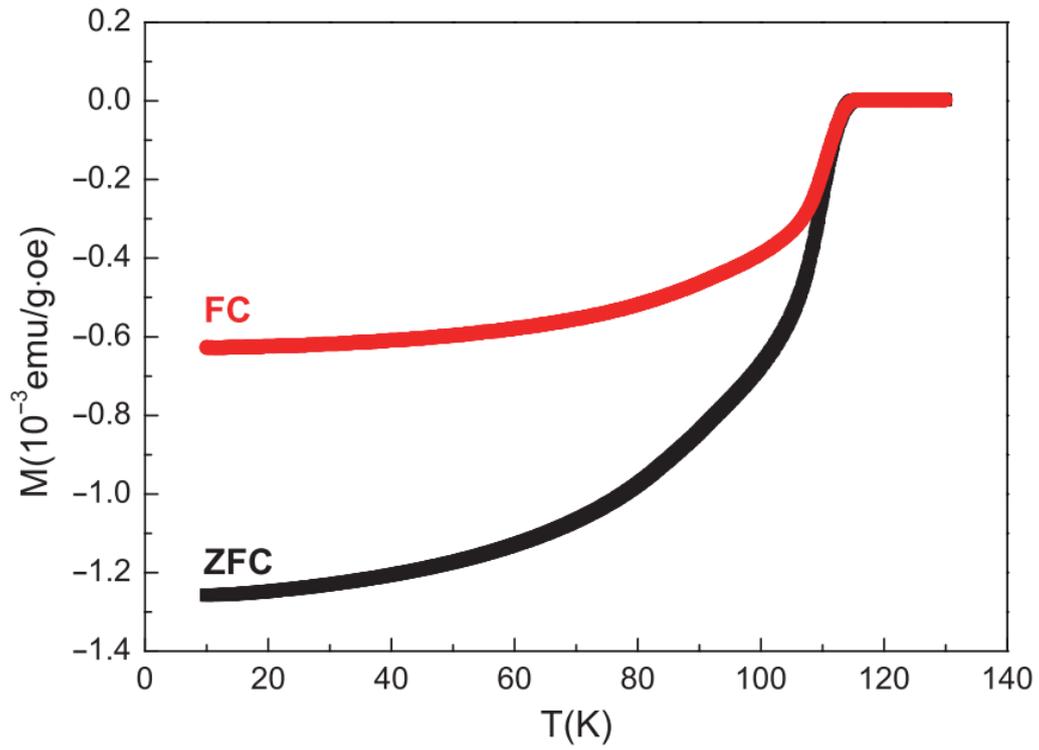

**Figure 4** (Color online) Temperature dependence of magnetic susceptibility of S2 measured in ZFC and FC modes with an applied field of 10Oe. The transition temperature is about 116 K which may correspond to Ga-1234 or Ga-1223 phases.